\documentclass[12pt,twoside]{article}
\usepackage{latexsym}
\usepackage{longtable}
\usepackage{epsfig}
\usepackage{a4wide}
\usepackage{amsmath,amsthm,amsfonts,amssymb,bbm}
\usepackage{graphicx,psfrag}
\usepackage{cite}
\numberwithin{equation}{section} 
\usepackage{epstopdf}
\graphicspath{{images/}}

\begin{document}
\begin{titlepage}
\begin{center}
\vspace*{2cm} {\Large {\bf Exact Height Distributions for the
KPZ Equation \medskip\\with Narrow Wedge Initial
Condition\bigskip\bigskip\\}} {\large{Tomohiro Sasamoto$^\star$ and
Herbert Spohn$^\dag$}}\bigskip\bigskip\\
  {$^\star$
  Department of Mathematics and Informatics,
 Chiba University,\\ 1-33 Yayoi-cho, Inage, Chiba 263-8522, Japan\\
 e-mail:~{\tt sasamoto@math.s.chiba-u.ac.jp\medskip}}\\
{$^\dag$  Zentrum Mathematik and Physik Department, TU M\"unchen,\\
 D-85747 Garching, Germany\\e-mail:~{\tt spohn@ma.tum.de}}\\

\end{center}
\vspace{5cm} \textbf{Abstract.} We consider the KPZ
equation in one space dimension with narrow wedge initial
condition, $h(x,t=0)=- |x|/\delta$, $\delta\ll 1$. Based on previous
results for the weakly asymmetric simple exclusion process with 
step initial conditions, we obtain a determinantal
formula for the one-point distribution of the solution $h(x,t)$
valid for any $x$ and $t>0$. The corresponding distribution function
converges in the long time limit, $t\to\infty$, to the Tracy-Widom
distribution. The first order correction is a shift of order
$t^{-1/3}$.  We provide numerical computations based on
the exact formula.

\end{titlepage}

\section{Introduction}

In a seminal paper \cite{KPZ} Kardar, Parisi, and Zhang (KPZ)
proposed a continuum equation for surface growth, which in the
particular case of one space-dimension reads
\begin{equation}\label{1.1}
\frac{\partial}{\partial
t}h=\tfrac{1}{2}\lambda\big(\frac{\partial}{\partial x}h\big)^2
+\nu\frac{\partial^2}{\partial x^2}h+\sqrt{D}\eta\,.
\end{equation}
Here $h(x,t)$ is the height profile at time $t$, $t\geq 0$, and $x\in
\mathbb{R}$. $\lambda$ is the strength of the nonlinearity.
(\ref{1.1}) is invariant under
$\lambda\rightsquigarrow -\lambda$ and $h\rightsquigarrow-h$. For
later convenience we choose $\lambda>0$. $\nu>0$ is the coefficient
of diffusive relaxation, $\eta$ normalized Gaussian white noise with
covariance $\langle\eta(x,t)\eta(x',t')\rangle=\delta(x-x')
\delta(t-t')$, and $\sqrt{D}$ the noise strength. (\ref{1.1}) is the
KPZ equation in one space dimension.

Typical realizations of the white noise $\eta(x,t)$ are far from regular functions. 
The solution $h(x,t)$ partially  inherits this roughness of the noise and therefore the square
$(\partial h/\partial x)^2$ is ill-defined, in general.
Nevertheless meaningful solutions
can be constructed by suitable approximation schemes. They will be
explained in detail below. The most direct one can be easily stated.
One smoothens $\eta$ to $\eta_\kappa$ as
\begin{equation}\label{1.1a}
\eta_\kappa (x,t)=\int \mathrm{d}x'
\kappa \varphi(\kappa(x-x')) \eta(x',t) = \varphi_\kappa\ast \eta(x,t)
\end{equation}
 with $\varphi$ some smooth, localized,
and normalized smearing function. Then $\eta_\kappa\to \eta$ as
$\kappa\to\infty$. (\ref{1.1}) with noise $\eta_\kappa$ has
well-defined solutions, denoted by $h_\kappa(x,t)$. They move with a uniform
background velocity $v_\kappa$ along the $h$-direction. $v_\kappa\to
\infty$ as $\kappa\to\infty$, but $h_\kappa(x,t)-v_\kappa t$ has a
limit. Since $v_\kappa$ sets merely the choice of a reference frame,
the claim is that under this limit procedure the fluctuation
properties remain intact.

While meaningful solutions are thus ensured, very little is known
about their properties. To make $\partial h(x,t)/\partial x$
stationary, one has to start the KPZ equation with two-sided
Brownian motion. With this input, one argues that the height
fluctuations will grow as $t^{1/3}$, while the transverse correlation
length increases as $t^{2/3}$. Very recently it has been
proved that the variance of the stationary two-point
function increases as $t^{4/3}$ by providing suitable upper and
lower bounds  \cite{BQS}.

In our investigation we choose as initial data the narrow wedge
\begin{equation}\label{1.2}
h(x,0)= -|x|/\delta
\end{equation}
with $\delta \ll 1$. This may look artificial. However, for short times the nonlinearity dominates
and, ignoring the other terms in the equation, 
$h$ spreads very rapidly into the parabolic profile
\begin{equation}\label{1.3}
h(x,t)=
\begin{cases}
-x^2/2\lambda t & \textrm{for }|x|\leq 2\lambda t/\delta\,, \\
-|x|/\delta & \textrm{for } |x|>2\lambda t/\delta\,,
\end{cases}
\end{equation}
which should be viewed as the top part of a growing droplet.
Physically one thereby  covers the case of macroscopically curved
height profiles.

We will provide a determinantal formula for the one-point distribution
of $h(x,t)$ at prescribed $x$ and valid for all  $t>0$, which is
exact in the properly normalized limit $\delta\to 0$. The formula
will be given in Equation (\ref{4.33}). Numerical plots based on this
formula are provided in Section 6. 

The exact solution is constructed from  the corner growth model,
which we introduce first. The corner growth is
the stochastic evolution of a height function $h(j,t)$,
$j\in\mathbb{Z}$, $t\geq 0$, which takes integer values constrained
by $|h(j+1,t)-h(j,t)|=1$. The initial condition is the wedge
$h(j,0)=-|j|$. In a short time interval $\mathrm{d}t$, independently for each
height, at a local minimum the height function increases by 2 with
probability $p\mathrm{d}t$ and stays put with probability $1-p\mathrm{d}t$.
Correspondingly, at a local maximum it decreases by 2 with probability
$q\mathrm{d}t$ and stays put with probability $1-q\mathrm{d}t$. These rules
respect the constraint. We set $q+p=1$ and $0\leq p < q \leq 1$,
so the height will decrease on average. It has been proved in
\cite{TW,TW2}, see also \cite{J} for the case $p=0$, that
\begin{equation}\label{1.4}
h(0,t) \cong -\tfrac{1}{2}(q-p)t + 2^{-1/3}((q-p)t)^{1/3} \xi_\mathrm{TW}
\end{equation}
for large $t$. $1/3$ is the scaling exponent predicted already on the basis of the
KPZ equation (\ref{1.1}). $\xi_\mathrm{TW}$ is a random amplitude
which is Tracy-Widom distributed, \textit{i.e.}
\begin{equation}\label{1.5}
\mathbb{P}(\xi_\mathrm{TW}\leq s)= F_\mathrm{TW}(s) = \det (1-P_s K_\mathrm{Ai}P_s)
\end{equation}
with $K_\mathrm{Ai}$ the Airy kernel,
\begin{equation}\label{1.6}
K_\mathrm{Ai}(x,y)=\int^\infty_0 \mathrm{d}\lambda
\mathrm{Ai}(x+\lambda)\mathrm{Ai}(y+\lambda)\,,
\end{equation}
$\mathrm{Ai}$ the Airy function, and with $P_s$ the projection onto the interval $[s,\infty)$.
$P_sK_\mathrm{Ai}P_s$ is a trace class operator on $L^2(\mathbb{R})$.
Hence the Fredholm determinant (\ref{1.5}) is well-defined on $L^2(\mathbb{R})$.

As will be discussed in detail, the exact KPZ solution is obtained
through the corner growth model in the limit of small asymmetry,
more precisely one requires $q-p\ll 1$, space of order $(q-p)^{-2}$,
and time of order $(q-p)^{-4}$. 
For large $t$ the exact solution
attains the Tracy-Widom statistics as in (\ref{1.4}). The solution also
predicts that the first order correction to Tracy-Widom is a shift
of order $t^{-1/3}$. Such qualitative features
remain valid over a much wider range of $q-p$ than anticipated by the method of construction
\cite{SS3}.
For the corner growth
model the maximal asymmetry is $q=1$. We expect that upon decreasing
$q$ towards $1/2$ finer and finer properties of the exact solution are revealed.
At present such claims can be tested only through Monte Carlo
simulations. For example, at $q=1$ for $t=10^3$ MC steps 
the average height is $- 512$ and its distribution 
has an effective range of 30 steps each of size
2. Even at this discrete level the Tracy-Widom distribution is an
accurate approximation provided one includes a shift of 2.9 
to the right relative to the approximation of (\ref{1.4}). For smaller values of $q$, approximately at $q=0.78$,
the prefactor of the $t^{-1/3}$ shift changes sign, which again can
be understood from the KPZ solution. To leading order in $t$ the solution
is shifted by $\gamma_t^{-1}(-0.577
+2\log(q-p))$, $\gamma_t = 2^{-1/3}(q-p)^{4/3}t
^{1/3}$, which is indeed negative. To study finer details would
require longer Monte Carlo runs, which have not been carried out
yet. Available is the $t^{-1/3}$ correction of the Tracy-Widom
distribution for the polynuclear growth model, as based on an exact recursion
formula \cite{Pra1}. For large
$t$ a definite limit is approached. The correction has a single
node. Thus it agrees qualitatively  with the $t^{-1/3}$ correction of
our solution, which is proportional to $F''_\mathrm{TW}(s)$, but the precise shape of the correction is different. Apparently
the first order correction to Tracy-Widom
depends on the details of the model.


\section{The Cole-Hopf solution} \label{sec2}
\setcounter{equation}{0}

We start from the mollified KPZ equation
\begin{equation}\label{2.1}
\frac{\partial}{\partial
t}h_\kappa=\tfrac{1}{2}\lambda(\frac{\partial}{\partial
x}h_\kappa)^2 +\nu\frac{\partial^2}{\partial
x^2}h_\kappa+\sqrt{D}\eta_\kappa\,,
\end{equation}
where the Gaussian noise $\eta_\kappa$ is smeared as in (\ref{1.1a}) and hence 
has the covariance
\begin{equation}\label{2.1a}
\langle\eta_\kappa(x,t)\eta_\kappa(x',t')\rangle=\varphi_\kappa\ast
\varphi_\kappa(x-x') \delta(t-t')\,.
\end{equation}
The mollifier $\varphi_\kappa$ satisfies $\varphi_\kappa(x)=\kappa \varphi(\kappa
x)$, $\varphi\geq 0$, $\varphi$ of rapid decay, $\varphi(x)=\varphi(-x)$, and $\int
\varphi(x)\mathrm{d}x=1$. The average over space-time white noise is
denoted by $\langle\cdot\rangle$. (\ref{2.1}) becomes linear through
the Cole-Hopf transformation
\begin{equation}\label{2.2}
Z_\kappa(x,t)=\exp [(\lambda/2\nu) h_\kappa(x,t)]\,,
\end{equation}
in which case $Z_\kappa$ satisfies
\begin{equation}\label{2.3}
\frac{\partial}{\partial t} Z_\kappa = \nu\frac{\partial^2}{\partial
x^2}Z_\kappa +(\lambda\sqrt{D}/2\nu)\eta_\kappa Z_\kappa\,.
\end{equation}
The solution to (\ref{2.3}) can be written in terms of the
Feynman-Kac formula by introducing an auxiliary  standard Brownian motion
$b(t)$ with variance $t$. Denoting $\mathbb{E}_x$ its expectation when
starting at $x$, one has
\begin{equation}\label{2.4}
Z_\kappa(x,t)=\mathbb{E}_x\Big(\exp\big[(\lambda\sqrt{D}/2\nu)\int^t_0
\eta_\kappa (b(2\nu s),t-s)\mathrm{d}s\big]
Z(b(2\nu t),0)\Big)
\end{equation}
for some deterministic initial condition $Z(\cdot,0)$.

In our context, the initial condition is a narrow wedge of the form (\ref{1.3}) with a global off-set. More precisely we 
assume
$h(x,0)=(2\nu/\lambda)(-\delta^{-1}|x| -\log2\delta)$, $\delta\ll 1$. Then the Cole-Hopf
transform $Z(x,0) = (2\delta)^{-1}\exp[-|x|/\delta]$ which is sharply peaked at $x=0$ and, as an
idealization, we will set
\begin{equation}\label{2.5}
Z(x,0)=\delta(x)\,.
\end{equation}

Note that the average
\begin{equation}\label{2.6}
\langle
Z_\kappa(x,t)\rangle=\exp\big[\tfrac{1}{2}(\lambda\sqrt{D}/2\nu)^2
 \varphi_\kappa \ast \varphi_\kappa(0)t\big]\mathbb{E}_x
 \big(Z(b(2\nu t),0)\big)\,.
\end{equation}
Since $\varphi_\kappa \ast \varphi_\kappa(0) \simeq \kappa$, it diverges as $\exp[c\kappa]$
with a suitable constant $c > 0$. Thus to have a limit as
$\kappa\to\infty$, one better divides the right hand side of
(\ref{2.4}) by $\langle Z_\kappa(x,t)\rangle$, which
corresponds to the Wick ordering of the exponential, see for example \cite{S78}. Hence we define
\begin{equation}\label{2.7}
{:\!Z_\kappa(x,t)\!:}= Z_\kappa(x,t)
\exp\big[ -\tfrac{1}{2}(\lambda\sqrt{D}/2\nu)^2 \varphi_\kappa \ast \varphi_\kappa(0)t\big]\,.
\end{equation}
It is proved in \cite{BG} that, for continuous $Z(\cdot,0)$ with an
explicit decay condition for $|x|\to\infty$, the limit
\begin{equation}\label{2.8}
\lim_{\kappa\to\infty} {:\!Z_\kappa (x,t)\!:}= Z(x,t)
\end{equation}
exists. $Z(x,t)$ is a well-defined stochastic process. With
probability one it has sample paths which are continuous in both
$x,t$. In addition ${Z(x,t)} > 0$ provided $Z(x,0)\geq 0$ and $Z(\cdot,0) \neq 0$. With
this input we \textit{define}
\begin{equation}\label{2.9}
h(x,t)=\frac{2\nu}{\lambda}\log Z(x,t)
\end{equation}
as the solution of the KPZ equation with $\delta$-correlated noise. For later purposes it is convenient
to introduce,  for arbitrary $\alpha > 0$, 
\begin{equation}\label{4.26a}
\alpha h_{\alpha}(x,t)=\log\mathbb{E}_x\Big({:\!\exp\big[\alpha
\int^t_0 \mathrm{d}s \eta(b(s),t-s)\big]\!\!:}\, \delta(b(t))\Big)\,.
\end{equation}
Clearly $h_\alpha$ is the solution to the KPZ equation with $D=1$ and $\alpha = \lambda/2\nu$.
By construction the first exponential moment is centered as
\begin{equation}\label{4.27}
\langle\mathrm{e}^{\alpha h_{\alpha}(x,t)}\rangle = (2\pi
t)^{-1/2} \mathrm{e}^{-(x^2/2t)}\,.
\end{equation}

From the point of view of the height function, the subtraction
corresponds to
\begin{equation}\label{2.10}
h_\kappa(x,t)-\tfrac{1}{2}(\lambda D/2\nu) \varphi_\kappa\ast
\varphi_\kappa (0)t\,.
\end{equation}
Hence the uniform velocity, alluded to in the introduction, is given by
\begin{equation}\label{2.10a}
v_\kappa=\tfrac{1}{2}(\lambda D/2\nu) \varphi_\kappa\ast \varphi_\kappa
(0)\,.
\end{equation}

(\ref{2.4}) can be viewed also as a directed polymer subject to a
space-time white noise random potential. With the initial condition
(\ref{2.5}) the polymer $b(s)$ starts at $x$ and
ends at 0 at time $2\nu t$, which corresponds to a point-to-point directed polymer.
The potential of the random environment is somewhat singular, hence
the subtraction. $Z(x,t)$ is the random partition function
of the directed polymer at the ``inverse temperature''
$\lambda\sqrt{D}/2\nu$. From the perspective of disordered systems,
of interest is the random free energy
$(2\nu/\lambda)\log Z(x,t)$, which is nothing else than the
KPZ height $h(x,t)$.

The observation leading to (\ref{2.6}) extends to higher moments. We
consider the $n$-th moment of ${:\!Z_\kappa(t)\!:}$ and for this purpose
introduce $n$ independent standard Brownian motions $b_j(t)$,
$j=1,\ldots,n$, as replicas of $b(t)$. Then
\begin{eqnarray}\label{2.11}
&&\hspace{-6pt}\langle({:\!Z_\kappa (x,t)\!:})^n\rangle=\mathbb{E}_x
\times\ldots \times \mathbb{E}_x\Big(\big\langle
\exp\big[(\lambda\sqrt{D}/2\nu)\sum^n_{j=1}
\int^t_0\eta_\kappa(b_j(2\nu s),t-s)
\mathrm{d}s\nonumber\\
&&\hspace{120pt}-
\tfrac{1}{2}n(\lambda\sqrt{D}/2\nu)^2 \varphi_\kappa\ast \varphi_\kappa
(0)t\big]\big\rangle \prod^n_{j=1} Z(b_j(2\nu t),0)\Big)\nonumber\\
&&\hspace{-6pt}=\mathbb{E}_x \times\ldots \times
\mathbb{E}_x\Big(\exp\big[\tfrac{1}{2}(\lambda\sqrt{D}/2\nu)^2
\sum^n_{i,j=1,i\neq j} \int^t_0 \varphi_\kappa\ast \varphi_\kappa
(b_i(2\nu s)-b_j(2\nu s))\mathrm{d}s\big]\nonumber\\
&&\hspace{120pt}\times\prod^n_{j=1} Z(b_j(2\nu t),0)\Big)\,,
\end{eqnarray}
which is the path integral for $n$ quantum particles with pair potential
$-\varphi_\kappa\ast \varphi_\kappa$. Thus the Wick ordering of the exponential in 
(\ref{2.8}) is identical to the normal ordering
of the pair interaction. $\varphi_\kappa(x)\to \delta(x)$ as $\kappa\to\infty$ and (\ref{2.11})
converges to the path integral of a system of $n$ quantum
particles on the line interacting through an attractive
$\delta$-potential. The corresponding hamiltonian reads
\begin{equation}\label{2.12}
H_n=-\sum^n_{j=1} \frac{1}{2}\frac{\partial^2}{\partial
x^2_j}-\frac{1}{2}\alpha^2\sum^n_{i,j=1,i\neq j}\delta (x_i-x_j)\,,
\end{equation}
where
\begin{equation}\label{2.12a}
\alpha=(2\nu)^{-3/2}\lambda D^{1/2}\,.
\end{equation}
 Since $\mathrm{e}^{-t 2\nu H_n}$ has a continuous kernel, it is
 a meaningful limit to
set $Z(x,0)=\delta(x)$ in (\ref{2.11}).  Let us denote the $n$-particle state
$\prod^n_{j=1}\delta(x-x_j)$ by $|x\rangle$. Then
\begin{equation}\label{2.14}
\langle Z(x,t)^n\rangle=\langle 0|\mathrm{e}^{-t 2\nu
H_n}|x\rangle
\end{equation}
with inner product in $L^2(\mathbb{R}^n)$.

For later purposes we record that, for the initial condition
(\ref{2.5}),
\begin{equation}\label{2.14a}
\langle{:\!Z_\kappa (x,t/2\nu)\!:}\rangle=\frac{1}{\sqrt{2\pi
t}}\exp[-x^2/2 t]
\end{equation}
and hence also in the limit $\kappa\to\infty$, in agreement with
(\ref{2.14}) for $n=1$.

Very recently, Bala\'{z}s, Quastel, and Sepp\"{a}l\"{a}inen
\cite{BQS} studied the KPZ equation with initial condition
$Z(x,0)=\exp[(\lambda/2\nu)\tilde{b}(x)]$, where $\tilde{b}(x)$ is two-sided
Brownian motion. They consider the solution as defined in
(\ref{2.9}) and prove, amongst other related results, upper and
lower bounds of the form
\begin{equation}\label{2.15}
c_- t^{2/3}\leq \mathrm{Var}\big(h(x,t)\big)\leq c_+ t^{2/3}\,.
\end{equation}
We cannot treat their case, because we miss the yet to be
accomplished extension of the contour integration formula in
\cite{TW} to initial conditions given by the Bernoulli $\frac{1}{2}$
measure, see \cite{TW4} for recent progress.


\section{The WASEP limit} \label{sec3}
\setcounter{equation}{0}

Bertini and Giacomin \cite{BG} (for short BG) discovered a rather different, and in
a way physically better motivated, limiting procedure for the
construction of solutions to the KPZ equation. They consider the
weakly asymmetric version of the simple exclusion process (WASEP),
which is equivalent to the corner growth model alluded to already in the introduction.
In the height picture this corresponds to the following stochastic evolution: A
height function $h:\mathbb{Z}\to\mathbb{Z}$ has to satisfy the
constraint
\begin{equation}\label{3.1}
|h(j+1)-h(j)|=1
\end{equation}
for all $j\in\mathbb{Z}$. Independently, each height variable waits
a unit exponentially distributed time. Then $h(j)\rightsquigarrow
h(j)-2$ with probability $q$ and $h(j)\rightsquigarrow h(j)+2$ with
probability $p$, $p+q=1$. Either transition is suppressed  in
case (\ref{3.1}) is violated. The height differences are governed by the
partially asymmetric simple exclusion process (PASEP). Here, because of
(\ref{3.1}), there is at most one particle per site and, under the  exclusion
constraint, particles jump with rate $p$ to the
right and rate $q$ to the left. We choose $q>p$. Thus $h$ decreases to
$-\infty$.

Weak asymmetry means that $q-p$ is small. The precise
asymptotics can be studied in a scaling limit with dimensionless
scale parameter $\varepsilon$, $\varepsilon\ll 1$. We choose a
diffusive space-time scaling, \textit{i.e.} time as
$\varepsilon^{-2} t$, $t=\mathcal{O}(1)$, space as $\lfloor
\varepsilon^{-1}x\rfloor$, $x\in\mathbb{R}$, $\lfloor\cdot\rfloor$
denoting integer part. The choice $q-p=\mathcal{O}(\varepsilon)$ is
the widely studied WASEP, which has the deterministic KPZ equation
$(D=0)$ as limit and Gaussian fluctuations relative to that profile
\cite{DM}. In contrast to the conventional choice BG consider the
crossover scale with an asymmetry equal to $\sqrt{\varepsilon}$,
see \cite{SS1} for a more complete discussion. For purpose of
comparison with the KPZ equation we set more generally
\begin{equation}\label{3.1a}
    q-p=\beta\sqrt{\varepsilon}\,, \quad \beta > 0\,.
\end{equation}
For the initial height profile BG assume
\begin{equation}\label{3.2}
h(j,0)=\lfloor \varepsilon^{-1/2} \phi(\varepsilon j)\rfloor \,,
\end{equation}
where $\phi$ is a smooth function with some decay condition at
infinity. Here $\lfloor\cdot\rfloor$ means integer part subject to (\ref{3.1}).
We denote by $h^\varepsilon(j,t)$ the WASEP time-evolved
height profile, the superscript $\varepsilon$ reminding that the
rates are $\varepsilon$-dependent. BG prove the limit
\begin{equation}\label{3.3}
\lim_{\varepsilon\to 0} \sqrt{\varepsilon}\big(h^\varepsilon(\lfloor
\varepsilon^{-1} x\rfloor, \varepsilon^{-2}t)+\tfrac{1}{2}
t\varepsilon^{-3/2}\big)=h(x,t)+\tfrac{1}{24} t\,,
\end{equation}
where $h(x,t)$ is the solution of the KPZ equation, as defined in the
previous section, with initial condition $\phi$ and parameters
$\lambda= 1$, $\nu=\frac{1}{2}$, $D=1$.

Note that $h(j+1,0)-h(j,0)=\mathcal{O}(\sqrt{\varepsilon})$ and the
particle density hence  deviates by order $\sqrt{\varepsilon}$ from
$\frac{1}{2}$. At that density the average time integrated particle
current equals $- t\varepsilon^{-3/2 }/4$ to leading order, which is
the subtraction on the left of (\ref{3.3}). (There is a factor 2
when switching from integrated currents to heights.) The subtraction
on the right is more subtle. As will be discussed in more detail
below, it originates from the normalization condition
\begin{equation}\label{3.3a}
\langle \mathrm{e}^{h(x,t)} \rangle= \int (2\pi t)^{-1/2} \exp
[-(x-x')^2/2t] \mathrm{e}^{\phi(x')} \mathrm{d}x'\,.
\end{equation}

To implement for the WASEP height the narrow wedge initial
condition from Section 2, the natural choice would be density
$\frac{1}{2}-\delta^{-1}\sqrt{\varepsilon}$ to the left and density
$\frac{1}{2}+\delta^{-1}\sqrt{\varepsilon}$ to the right of the
origin. Unfortunately, for such initial conditions there are no
manageable formulas available and we have to take resort to initial
conditions which no longer are covered by (\ref{3.3}) and the
results in \cite{BG}.

As standing assumption, the WASEP has asymmetry
$\beta\sqrt{\varepsilon}$ and an initial configuration such that all
sites from 1 to $\infty$ are filled and all other sites are empty.
The corresponding initial height profile is $h(j,0)=-|j|$. After some
short initial time span, the typical height profile becomes
\begin{equation}\label{3.4}
h^\varepsilon(j,\varepsilon^{-2}t)\cong
\begin{cases}
-\frac{1}{2}(j^2/\beta t) \varepsilon^{3/2}
- \frac{1}{2}\beta t\varepsilon^{-3/2} & \textrm{for }|j|\leq \beta t\varepsilon^{-3/2}\,, \\
-|j| & \textrm{for } |j|>\beta t\varepsilon^{-3/2}\,,
\end{cases}
\end{equation}
compare with (\ref{1.3}). If one restricts $j$ to the interval $[-\varepsilon^{-1},
\varepsilon^{-1}]$, then there one has a parabolic profile, which
scales exactly as assumed in (\ref{3.2}) and thus supports that
(\ref{3.3}) remains valid for the initial sharp wedge. In \cite{SS1} we obtained the one-point
height distribution for the WASEP with 0\,-1 initial condition in the
limit $\varepsilon\to 0$. Thus the plan is to use this result in the
construction of the one-point distribution of the KPZ equation with
narrow wedge initial condition.

Let us denote by $\eta_j(t)=0,1$ the occupation variables of the WASEP
at site $j$ and time $t$. Since there is a random index $j_0$ such
that $\eta_j(t)=0$ for $j<j_0$, one can define the WASEP height by
\begin{equation}\label{4.1}
h^\varepsilon(j,t)= -2\sum^j_{\ell=-\infty}\eta_\ell(t)+j\,.
\end{equation}
Let $\mathcal{J}^\varepsilon(j,t)$ be the particle current across the bond
$(j,j+1)$ integrated up to time $t$. Then
\begin{equation}\label{4.1a}
h^\varepsilon(j,t)=2\mathcal{J}^\varepsilon(j,t)-|j|\,.
\end{equation}
In \cite{SS1} we studied the distribution of $\mathcal{J}^\varepsilon$ at
$j=\lfloor y\beta t \varepsilon^{-3/2} +\varepsilon^{-1} x\rfloor$
and time $\varepsilon^{-2}t$. To identify the height statistics of the KPZ
equation it suffices to take $y=0$, in which case
\begin{eqnarray}\label{4.2}
&&\hspace{-42pt}\lim_{\varepsilon\to
0}\mathbb{P}\big(\sqrt{\varepsilon}\beta h^{\varepsilon} (\lfloor
\varepsilon^{-1}x\rfloor,\varepsilon^{-2}t) +\tfrac{1}{2} \beta^2 t
\varepsilon^{-1}+(x^2/2t) -\log (2\beta\sqrt{\varepsilon})\leq
 \gamma_t s\big)\nonumber\\
&&\hspace{107pt}=F_t(s)\,,
\end{eqnarray}
with
\begin{equation}\label{4.2a}
\gamma_t=2^{-1/3} (\beta^4 t)^{1/3}\,.
\end{equation}

$F_t(s)$ is a $t$-dependent family of distribution functions. To
define them, let us introduce the Airy
kernel $K_\mathrm{Ai}(x,y)$ of (\ref{1.6}),
the kernel
\begin{eqnarray}\label{4.4}
&&\hspace{-42pt} B_t(x,y) = K_\mathrm{Ai}(x,y)\nonumber\\
&&\hspace{-10pt} +\int^\infty_0
\mathrm{d}\lambda(\mathrm{e}^{\gamma_t\lambda}
-1)^{-1}\big(\mathrm{Ai}(x+\lambda)\mathrm{
Ai}(y+\lambda)-\mathrm{Ai}(x-\lambda)\mathrm{ Ai}(y-\lambda)\big)\,,
\end{eqnarray}
and the unnormalized projection $P_\mathrm{Ai}$ with kernel
\begin{equation}\label{4.5}
P_\mathrm{Ai}(x,y)=\mathrm{Ai}(x)\mathrm{Ai}(y)\,.
\end{equation}
Then
\begin{eqnarray}\label{4.6}
&&\hspace{-42pt}F_t(s)= 1- \int^\infty_{-\infty} \mathrm{d}u \exp[-
\mathrm{e}^{\gamma_t(s-u)}]
\nonumber\\
&&\hspace{60pt}\times \big(\det (1-P_u(B_t-P_\mathrm{Ai})P_u) -\det(1-P_u B_t P_u)\big)\,.
\end{eqnarray}
Here $P_u$ projects onto $[u,\infty)$. $P_u B_t P_u$ and $P_u
P_\mathrm{Ai} P_u$ are trace class
operators in $L^2(\mathbb{R})$. Hence the Fredholm determinants in
(\ref{4.6}) are well-defined.

Note that the probability density $\frac{d}{ds}F_t(s)$ is the
convolution of the Gumbel probability density
$\gamma_t\mathrm{e}^{\gamma_ts}\exp[-\mathrm{e}^{\gamma_t s}]$
and the difference of determinants,
\begin{equation}\label{4.7}
 g_t(u) = \det (1-P_u
(B_t-P_\mathrm{Ai})P_u)-\det (1-P_u B_t
P_u)\,.
\end{equation}
For $t\to\infty$ the Gumbel density converges to $\delta(s)$
and $g_t(u)$ to $F'_\mathrm{TW}(u)$, see  \cite{SS1}. Hence
\begin{equation}\label{4.7a}
\lim_{t\to\infty} F_t(s)=F_\mathrm{TW}(s)
\end{equation}
with $F_\mathrm{TW}$ the Tracy-Widom distribution function (\ref{1.5}).


\section{Centering and exact solution} \label{sec4}
\setcounter{equation}{0}

Through the limit to $\delta$-correlated noise we adjusted the first
moment of the partition function as
\begin{equation}\label{4.9}
\langle{Z(x,t/2\nu)}\rangle=\frac{1}{\sqrt{2\pi t}}\exp
\big[-\frac{x^2}{2t}\big]\,,
\end{equation}
which sets an important constraint for the WASEP limit. To exploit it, one notes that
a particular exponential moment of the WASEP also satisfies a closed
linear equation. This property was first proved by G\"{a}rtner \cite{G1,G2}, see also
\cite{BG} and the related contribution \cite{GS}. For the PASEP with fixed $p,q$ and
arbitrary initial conditions, let us denote by $\mathbb{E}_t$ the expectation with respect to
the height statistics at time $t$ and let us set
\begin{equation}\label{4.10}
f(j,t)=\mathbb{E}_t(\mathrm{e}^{\vartheta h(j)})\,.
\end{equation}
$\vartheta$ is adjusted  such that
\begin{equation}\label{4.11}
\mathrm{e}^{-2\vartheta}=\frac{p}{q}\,.
\end{equation}
Then $f$ is the solution of
\begin{eqnarray}\label{4.12}
&&\hspace{-22pt}\frac{d}{dt}f(j,t)=\frac{1}{\cosh\vartheta}\big(\tfrac{1}{2}
f(j+1,t)+\tfrac{1}{2} f(j-1,t)-f(j,t)\big)
+(\frac{1}{\cosh\vartheta}-1) f(j,t)\,, \nonumber\\[1ex]
&&\hspace{-10pt}f(j,0)=\mathrm{e}^{-\vartheta|j|}\,.
\end{eqnarray}
 (\ref{4.12}) can
be deduced by working out the time derivative as
\begin{equation}\label{4.13}
\frac{d}{dt}f(j,t)=\mathbb{E}_t(L \mathrm{e}^{\vartheta h(j)})
\end{equation}
with $L$ the generator of the Markov jump process for the heights.
To satisfy the constraint (\ref{4.9}) one has to consider the height
shifted by $\log f$ as
\begin{equation}\label{4.14}
\vartheta h^\varepsilon
(\lfloor\varepsilon^{-1}x\rfloor,\varepsilon^{-2}t) - \log
f(\lfloor\varepsilon^{-1}x\rfloor,\varepsilon^{-2}t)- (x^2/2 t)-\log\sqrt{2\pi  t}\,.
\end{equation}

$f$ can be written as Fourier integral,
\begin{eqnarray}\label{4.15}
&&\hspace{-18pt}f(\lfloor \varepsilon^{-1}x\rfloor,\varepsilon^{-2}t)\nonumber\\
&&\hspace{6pt}=\frac{1}{2\pi}\int^\pi_{-\pi} \mathrm{d}k \exp\big[(\cos
k-1)(\cosh \vartheta)^{-1}\varepsilon^{-2}t+ \mathrm{i}k\varepsilon^{-1}x\nonumber\\
&&\hspace{50pt}+ \log \widehat{f}(k,0)+\varepsilon^{-2}t ((\cosh
\vartheta)^{-1}-1)\big]
\end{eqnarray}
with the initial condition
\begin{equation}\label{4.16}
\widehat{f}(k,0)=(1-\mathrm{e}^{-2\vartheta})(1-2\mathrm{e}^{-\vartheta}
\cos k+\mathrm{e}^{-2\vartheta})^{-1}\,.
\end{equation}
$\log f(\lfloor\varepsilon^{-1}x\rfloor,\varepsilon^{-2}t)$ has to be computed including terms of
order 1. One has
\begin{equation}\label{4.17}
\vartheta=\beta\varepsilon^{1/2}+\tfrac{1}{3}\beta^3\varepsilon^{3/2}+
\mathcal{O}(\varepsilon^{5/2})\,,
\end{equation}
\begin{equation}\label{4.18}
(\cosh \vartheta)^{-1}=
1-\tfrac{1}{2}\beta^2\varepsilon-\tfrac{1}{8}\beta^4\varepsilon^2+
\mathcal{O}(\varepsilon^3)\,.
\end{equation}
The initial condition (\ref{4.16}) is approximated by
$2/\beta\sqrt{\varepsilon}$. Then the Fourier integral is the
transition probability of a simple random walk with rate $(\cosh
\vartheta)^{-1}$ starting at 0, which at the given scale can be
approximated by a normalized Gaussian. Therefore
\begin{eqnarray}\label{4.20}
&&\hspace{-38pt}-\log f(\lfloor \varepsilon^{-1}x\rfloor,\varepsilon^{-2}t)\nonumber\\[1ex]
&&\hspace{-18pt}=\tfrac{1}{2}\beta^2 t\varepsilon^{-1}+
(x^2/2t)+\tfrac{1}{8}\beta^4 t +\log(\sqrt{2\pi
t}/\varepsilon)+\log(\beta\sqrt{\varepsilon}/2)+\mathcal{O}(\sqrt{\varepsilon})\,.
\end{eqnarray}

Inserting in (\ref{4.14}) and taking the deterministic approximation
to $h^\varepsilon$ for the second order in the expansion of $\vartheta$, one
arrives at
\begin{eqnarray}\label{4.21}
&&\hspace{-38pt}\vartheta h^\varepsilon(\lfloor
\varepsilon^{-1}x\rfloor,\varepsilon^{-2}t) -\log f(\lfloor \varepsilon^{-1}x\rfloor,
\varepsilon^{-2}t)- (x^2/2t)-\log\sqrt{2\pi t}\nonumber\\[1ex]
&&\hspace{-20pt}=\beta \sqrt{\varepsilon}h^\varepsilon(\lfloor
\varepsilon^{-1}x\rfloor,\varepsilon^{-2}t)+\tfrac{1}{2}\beta^2
t\varepsilon^{-1}-\tfrac{1}{24}\beta^4 t
- \log(2\sqrt{\varepsilon}/\beta)+\mathcal{O}(\sqrt{\varepsilon})\,.
\end{eqnarray}
Note that the $\varepsilon$-dependent part of the centering is
precisely as in (\ref{4.2}).

For the WASEP at the reference point $y=0$ the average particle
density $\rho$ equals $1/2$ and the noise strength equals
$\rho(1-\rho)=1/4$. Including the factor $2$ for the conversion between density and height
differences, the corresponding parameters for the KPZ equation are  $2\nu=1$ and
$D=1$. Since the asymmetry equals $\beta$, the WASEP height (\ref{4.21}) should hence
be compared with $\beta h_\beta(x,t)$, see (\ref{4.26a}).
Then,
stretching somewhat the results in \cite{BG},
for the WASEP limit it holds
\begin{equation}\label{4.28}
\lim_{\varepsilon\to 0}\sqrt{\varepsilon}\beta h^\varepsilon
(\lfloor \varepsilon^{-1}x\rfloor,\varepsilon^{-2}t)
+\tfrac{1}{2}\beta^2 t\varepsilon^{-1}-\tfrac{1}{24}\beta^4  t
-\log(2\sqrt{\varepsilon}/\beta) =\beta h_{\beta}(x,t) \,.
\end{equation}
The BG result (\ref{3.3}) corresponds to the particular case
$\beta=1$ and non-singular initial data.

Combining (\ref{4.28}) and (\ref{4.2}) one arrives at an exact
formula for the one-point distribution of $h_{\beta}$ which reads
\begin{equation}\label{4.29}
\mathbb{P}\big(\beta h_{\beta}(x,t)+\tfrac{1}{24}\beta^4 t
+(x^2/2t)-2\log\beta\leq \gamma_t s\big)= F_t(s) \,.
\end{equation}
To generalize to arbitrary $\lambda,\nu,D$ one starts from the definition
(\ref{2.9}) for the narrow wedge solution $h(x,t)$ of the KPZ equation and
uses the scale invariance of white noise and of
Brownian motion to obtain
\begin{eqnarray}\label{4.30}
&&\hspace{-18pt}(\lambda/2\nu)h(x,t/2\nu)=
\log\mathbb{E}_x\Big({:\!\exp\big[(\lambda\sqrt{D}/2\nu)
\int^{t/2\nu}_0 \eta(b(2\nu s),(t/2\nu)-s)\mathrm{d}s
\big]\!\!:}\delta(b(t))\Big)\nonumber\\
&&\hspace{71pt}=
\log\mathbb{E}_x\Big({:\!\exp\big[(\lambda\sqrt{D}/(2\nu)^{3/2})
\int^{t}_0 \eta(b(s),t-s)\mathrm{d}s \big]\!\!:}\delta(b(t))\Big)\nonumber\\
&&\hspace{71pt}=\alpha h_\alpha (x,t)
\end{eqnarray}
with $\alpha$ fixed to
\begin{equation}\label{4.32}
\alpha = (2\nu)^{-3/2} \lambda D^{1/2}\,.
\end{equation}

Inserting in (\ref{4.29}) one obtains the one-point distribution of
the KPZ equation with narrow wedge initial data,
\begin{equation}\label{4.33}
\mathbb{P}\big((\lambda/2\nu) h(x,t/2\nu)+\tfrac{1}{12}(\gamma_{
t})^3 +(x^2/2t)-2\log \alpha \leq \gamma_{t} s\big)=F_{
t}(s)\,,
\end{equation}
where $\gamma_t=2^{-1/3} (\alpha^4 t)^{1/3}$. The identity
(\ref{4.33}) is the central result of our contribution.
\medskip\\
\textit{Remark}: After posting this article it was brought  to our attention that G. Amir, I. Corwin, and J. Quastel
independently obtained the identity (\ref{4.33}) and the formula (\ref{4.6}) for the distribution function $F_t$.


\section{Some properties of the one-point distribution}
 \setcounter{equation}{0}
\textit{(i) Scale invariance.} In terms of the original parameters we define $\alpha$  as in (\ref{4.32})
and introduce dimensionless space,
time, and height through
\begin{equation}\label{5.a}
X=\alpha^2 x\,,\quad T=2\nu\alpha^4 t\,,\quad H=(\lambda/2\nu)h\,.
\end{equation}
Then the coefficients in the KPZ equation become $\lambda=1$,
$\nu=\frac{1}{2}$, and $D=1$, \textit{i.e.} $H(X,T)$ satisfies
\begin{equation}\label{5.b}
\frac{\partial}{\partial T}H=\frac{1}{2}\big(\frac{\partial H} {\partial
X}\big)^2 + \frac{1}{2}\frac{\partial^2}{\partial X^2}H+\eta\,.
\end{equation}
In these variables (\ref{4.33}) reads
\begin{equation}\label{5.c}
\mathbb{P}\big(H(X,T)+\tfrac{1}{24}T+(X^2/2T)-2\log \alpha\leq
\tilde{\gamma}_T s\big)=F_T(s)
\end{equation}
with $\tilde{\gamma}_T=2^{-1/3} T^{1/3}$. The distribution function $F_T(s)$ depends on $T$ 
through substituting $\gamma_t$ by $\tilde{\gamma}_T$. The solution (\ref{4.33})
satisfies the scale invariance of the KPZ equation, the term $2\log
\alpha$ resulting from the scaling of the initial condition
$Z(x,0) = \delta(x)$, see (\ref{4.26a}).\medskip\\
\textit{(ii) Stationarity.} Since for our initial conditions the
height profile is curved, no stationarity in $x$ is expected.
However, in the case of PNG droplet growth it was noted that height
fluctuations become stationary once the systematic curvature is
subtracted. The same feature appears for the KPZ equation.  Using
dimensionless units for simplicity,
we set
\begin{equation}\label{5.1}
h(x,t)=\log\mathbb{E}_0\Big({:\!\exp\big[
\int^t_0  \eta(b(s),s)\mathrm{d}s\big]\!\!:}\,\delta(b(t)-x)\Big)\,,
\end{equation}
where, with no modification of the statistics, the Brownian motion $b(s)$ has been time-reversed to
$b(t-s)$.
Let us
introduce the Brownian bridge $w(s)$ running from 0 at $s=0$ to $x$ at
$s=t$. In terms of Brownian motion,
\begin{equation}\label{5.4}
w(s)=b(s) -\frac{s}{t}(b(t)-x)\,.
\end{equation}
Denoting by $\mathbb{E}^{\mathrm{BB}}_{0,x}$ the expectation over a Brownian bridge
with endpoints $0,x$,
 we define
\begin{equation}\label{5.5}
\tilde{h}(x,t)=\log\mathbb{E}^{\mathrm{BB}}_{0,x}\big({:\!\exp\big[
\int^t_0  \eta(w(s),s)\mathrm{d}s\big]\!\!:}\big)\,,
\end{equation}
which satisfies the $x$-independent normalization
\begin{equation}\label{5.4a}
\langle\mathrm{e}^{\tilde{h}(x,t)}\rangle = 1\,.
\end{equation}
Using scale invariance, one checks that $x\mapsto
\tilde{h}(x,t)$ is stationary as a stochastic process for
fixed $t$.  On the level of the one-point distribution this property is
reflected by $F_t(s)$ being independent of $x$.
\medskip\\
\textit{(iii) Ground state energy of the $\delta$-Bose gas.} As
first shown by McGuire \cite{McG}, in units of (\ref{2.14}) with
$2\nu = 1$ the ground state energy of the $\delta$-Bose gas with $n$
particles is given by
\begin{equation}\label{5.9}
E_n=- \tfrac{1}{24}n(n^2-1)\alpha^4\,.
\end{equation}
By (\ref{2.14}),
\begin{equation}\label{5.10}
\langle Z(0,t)^n\rangle = \langle 0|\mathrm{e}^{-t
H_n}|0\rangle\cong \mathrm{e}^{-t E_n}
\end{equation}
for large $t$. Of course, there are subleading terms, which however
will be ignored at this stage. From the exact solution one has
\begin{equation}\label{5.11}
\mathbb{P}(\alpha h_{\alpha}(0,t)+\tfrac{1}{24}\alpha^4 t-2\log\alpha
\leq s)=F_t(s/\gamma_t)\,.
\end{equation}
Hence
\begin{eqnarray}\label{5.12}
&&\hspace{-24pt} \mathbb{E}\big(\exp\big[n\big(\alpha
h_{\alpha}(0,t)+\tfrac{1}{24}\alpha^4 t-2\log\alpha\big)\big]\big)\nonumber\\
&&\hspace{8pt}=\int \mathrm{d}s \int \mathrm{d}u \mathrm{e}^{ns}\mathrm{e}^{s}\mathrm{e}^{-\gamma_tu}
\exp \big[-\mathrm{e}^s \mathrm{e}^{-\gamma_t u} \big] g_t(u)\nonumber\\
&&\hspace{8pt}=n! \int \mathrm{d}u \mathrm{e}^{n \gamma_t u } g_t(u)
\end{eqnarray}
with $g_t(u)$ defined in (\ref{4.7}).

$\gamma_t$ diverges as $t^{1/3}$ for $t\to\infty$. Hence the
integral in (\ref{5.12}) is sensitive only to the right tail of $g_t(u)$.
Furthermore, by the argument before (\ref{4.7a}), $g_t(u)$ tends to $F'_{\mathrm{TW}}(u)$,
which has a right tail as $\exp[-\frac{4}{3}u^{3/2}]$, see e.g. \cite{PS}.
Therefore to leading order
\begin{equation}\label{5.13}
\int^\infty_0 \mathrm{d}u \exp[-\tfrac{4}{3}u^{3/2}+n \gamma_t
u]\cong \exp[ \tfrac{1}{24} \alpha^4t n^3]
\end{equation}
for $t\to\infty$. Combining with (\ref{5.12}),
\begin{equation}\label{5.14}
\lim_{t \to \infty} - \frac{1}{t} \log\mathbb{E}\big(\exp\big[n\alpha h_{\alpha}(0,t)\big]\big) =
-\tfrac{1}{24}n(n^2-1)\alpha^4 \,,
\end{equation}
in agreement with (\ref{5.9}), (\ref{5.10}).

As discussed in \cite{K2}, Chapters 9 and 10, the connection to the $\delta$-Bose gas
was used to support the scaling exponent $1/3$ and to study the right tail of the height distribution.
We added here the computation of the exact prefactor $4/3$ for the stretched exponential.
Note that the left hand tail of the Tracy-Widom density decays as $\exp[-|u|^3/12]$.
\section{Numerical evaluations}\label{sec6}
 \setcounter{equation}{0}

According to (\ref{4.33}) the one-point height statistics is
governed by the probability density
\begin{equation}\label{6.1}
\rho_t (s) =\frac{\mathrm{d}}{\mathrm{d}s}F_t(s) = \int^\infty_{-\infty}  \gamma_t
\mathrm{e}^{\gamma_t(s-u)}\exp \big[-\mathrm{e}^{\gamma_t(s-u)}\big]g_t(u)\mathrm{d}u\,,
\end{equation}
where $g_t(u)$ is defined in (\ref{4.7}). This expression is
sufficiently concrete to allow for some exploratory numerical
studies. In Appendix A we establish that $g_t(s)$ can be approximated for large $t$ as
\begin{equation}\label{6.2}
g_t(s)= \rho_\mathrm{TW}(s) +\widetilde{g}_t(s) +
\mathcal{O}(\gamma^{-6}_t)
\end{equation}
with the first order correction $\widetilde{g}_t(s)$ given in
(\ref{15}). Correspondingly one has the first order approximation to
$\rho_t(s)$ as
\begin{equation}\label{6.3}
\rho^{(1)}_t (s)= \int  \gamma_t \mathrm{e}^{\gamma_t(s-u)}\exp
\big[-\mathrm{e}^{\gamma_t(s-u)}\big]\big(\rho_\mathrm{TW}(u)+
\widetilde{g}_t(u)\big)\mathrm{d}u\,.
\end{equation}

$\widetilde{g}_t$ can be expressed through quantities related
to the Tracy-Widom distribution, see (\ref{26}), and tabulated in \cite{Pra}.
The evaluation is
displayed in Fig. 1, where the vertical axis corresponds to
$\gamma_t=1$.
In Fig. 2 we plot $\rho^{(1)}_t$
as defined in (\ref{6.3}).  $\rho^{(1)}_t$ turns
negative at $\gamma_t\cong 5$, which signals the border of validity of the first order approximation.
As shown in Fig. 3, at
$\gamma_t=2$ the difference
$\rho^{(1)}_t-\rho_\mathrm{TW}$ has two nodes while the difference
$\rho_t-\rho_\mathrm{TW}$ as displayed in Fig. 5 below has only a single
node, again indicating qualitative differences.

A more demanding approach is a numerical evaluation of the
determinants defining $g_t$. For this purpose the kernels of $B_t$
and $P_\mathrm{Ai}$ are computed using Mathematica. One chooses a
properly adjusted equally spaced grid of 120 points for each
variable resulting in a matrix $\{M_{ij}\}_{i,j= 1,\dots,120}$. The step size for $s$ is 0.1 covering the interval $[-6,6]$. 
The projection $P_s$ means to restrict $M$ to $i,j = (60+ 10s),\dots,120$. The respective determinant 
is then evaluated for various $s, t$ again using Mathematica. 
To reach higher precision and earlier times it would be
advisable to use the more refined methods of \cite{Bor}. 

$\rho_t$ is displayed for $\gamma_t=2,5,10$ in Fig. 4. The dotted line is the
limiting Tracy-Widom density. For better visibility, in Fig. 5 we
plot the difference $\rho_t(s)-\rho_\mathrm{TW}(s)$ at the same times.
In Fig. 6
we test directly the precision of the first order correction by
plotting $\widetilde{g}_t$ at $\gamma_t=10$ and $g_t - \rho_\mathrm{TW}(s)$.
While there is qualitative agreement we would have expected a much
smaller error, in particular at the left minimum. A much cruder
comparison is to plot merely the first moments of $\rho_t$ and
$\rho^{(1)}_t$. As displayed in Fig. 7 the agreement is fairly
precise beyond $\gamma_t=5$.

As argued in the Introduction, features of our exact solution should
be visible also in the corner growth model with finer and finer
details being revealed as $q$ is decreased from $1$ to
$\tfrac{1}{2}+\beta\sqrt{\varepsilon}$. To support this claim we
carried out a few Monte Carlo simulations. Time steps have a
sufficiently fine discretization and random sequential updating is
used. We measure the total number of signed jumps between sites 0 and 1.
 For $q=1$ at $t=10^3$ MC steps the height
distribution has an effective range of 30 equally spaced points.
Except for the shifted mean, the Tracy-Widom density is an accurate
approximation \cite{SS3}. For earlier times the height takes so few values that 
correction to $\rho_\mathrm{TW}$ does not seem to be a
meaningful notion. In Fig. 8 we plot the corresponding result for $q=0.6$ at 
$t=1024$ MC steps with  
an average over $10^4$ realizations. We compare with the exact solution at $\gamma_t = 0.94$.
Clearly, this is 
a better approximation than the Tracy-Widom distribution
indicated by the dashed line.\bigskip\bigskip\\
\textbf{Acknowledgements.} We are grateful to Michael Pr\"{a}hofer
for many illuminating discussions. H.~S. thanks Jeremy
Quastel for emphasizing the importance of the crossover WASEP.
This work is supported by a DFG grant. In addition T.S. acknowledges the support from
KAKENHI (9740044) and H.S. from Math-for-Industry
of Kyushu University.
\begin{appendix}
\section{Appendix: Expansion in $1/t$}\label{app.A}
\setcounter{equation}{0}
\textbf{Expansion of the determinants.}
We follow as closely as possible the notation in \cite{TW1}, simply
referred to as TW. As only difference we set
\begin{equation}\label{1}
    \psi(x)=\mathrm{Ai}(x)\quad \big(A(x)\textrm{ in TW}\big)
\end{equation}
and
\begin{equation}\label{2}
    \psi_\lambda(x)=\psi(x+\lambda)\,,\quad \psi_0(x)=\psi(x)\,.
\end{equation}
We define the operator
\begin{equation}\label{3}
B_t=K+C_t\,,
\end{equation}
with the kernels
\begin{equation}\label{3a}
K(x,y)=\int^\infty_0 \mathrm{d}\lambda
\psi_\lambda(x)\psi_\lambda(y)\,,
\end{equation}
formerly denoted by $K_{\mathrm{Ai}}$, and
\begin{equation}\label{5}
C_t(x,y)=\int^\infty_0 \mathrm{d}\lambda (\mathrm{e}^{\gamma_t
\lambda}-1)^{-1}
\big(\psi_\lambda(x)\psi_\lambda(y)-\psi_{-\lambda}(x)
\psi_{-\lambda}(y)\big)\,,
\end{equation}
and the unnormalized projection $P_\psi$ with kernel
$\psi(x)\psi(y)$.

We consider the difference
\begin{equation}\label{6}
g_t(s)=\det (1-K-C_t+P_\psi)-\det (1-K-C_t)\,.
\end{equation}
The projection $P_s$ in (\ref{4.7}) is taken into account by defining the determinant
in $L^2([s,\infty))$ with scalar product
$\langle\cdot,\cdot\rangle$. For fixed $s$ we want to expand
$g_t(s)$ in $\gamma^{-1}_t$, which means to regard $C_t$ as a small
perturbation.

One starts from
\begin{eqnarray}\label{7}
&&\hspace{-30pt}g_t=\det (1-K-C_t)\langle
\psi,(1-K-C_t)^{-1}\psi\rangle\nonumber\\[1ex]
&&\hspace{-18pt}=\det (1-K)\det(1-(1-K)^{-1}C_t)\langle
\psi,(1-K-C_t)^{-1}\psi\rangle\,,
\end{eqnarray}
expands the Fredholm determinant as
\begin{equation}\label{8}
\det(1-(1-K)^{-1}C_t)=1-\mathrm{tr}((1-K)^{-1}C_t)+\mathcal{O}(C^2_t)\,,
\end{equation}
and writes the resolvent as
\begin{equation}\label{9}
\langle \psi,(1-K-C_t)^{-1}\psi\rangle=\langle
\psi,(1-K)^{-1}\psi\rangle+\langle
\psi,(1-K)^{-1}C_t(1-K)^{-1}\psi\rangle+\mathcal{O}(C^2_t)\,.
\end{equation}
The leading term equals $F'/F$, see \cite{SS1}, where
\begin{equation}\label{8a}
F =  F_\mathrm{TW}(s) = \det (1-P_s K_\mathrm{Ai}P_s) = \det(1-K)\,.
\end{equation}
The
term linear in $C_t$ is
\begin{eqnarray}\label{10}
&&\hspace{-30pt}\widetilde{g}_t=F\big\{\langle
\psi,(1-K)^{-1}C_t(1-K)^{-1}\psi\rangle-\langle
\psi,(1-K)^{-1}\psi\rangle
\mathrm{tr}((1-K)^{-1}C_t)\big\}\nonumber\\[1ex]
&&\hspace{-18pt}=g_t-F'+\mathcal{O}(C^2_t)\,.
\end{eqnarray}
Inserting the definition of $C_t$ from (\ref{5}), one arrives at
\begin{equation}\label{11}
\widetilde{g}_t=F \int^\infty_0 \mathrm{d}\lambda
(\mathrm{e}^{\gamma_t \lambda}-1)^{-1}
\big(G(\lambda)-G(-\lambda)\big)
\end{equation}
with
\begin{equation}\label{12}
G(\lambda)= \langle
\psi,(1-K)^{-1}\psi_\lambda\rangle\langle
\psi_\lambda,(1-K)^{-1}\psi\rangle-\langle
\psi_\lambda,(1-K)^{-1}\psi_\lambda\rangle\langle
\psi,(1-K)^{-1}\psi\rangle\,.
\end{equation}
Rescaling by $\gamma_t$ one arrives at
\begin{equation}\label{13}
\widetilde{g}_t=F \gamma^{-1}_t\int^\infty_0 \mathrm{d}\lambda
(\mathrm{e}^\lambda-1)^{-1}
\big(G(\lambda/\gamma_t)-G(-\lambda/\gamma_t)\big)\,.
\end{equation}
Thus the expansion in $1/t$ amounts to a Taylor expansion of
$\widetilde{G}(\lambda)=G(\lambda)-G(-\lambda)$ at $\lambda=0$.

One has $\widetilde{G}(0)=0$, $\widetilde{G}''(0)=0$, since
$\widetilde{G}$ is odd. Furthermore $\widetilde{G}'(0)=0$ and
\begin{eqnarray}\label{14}
&&\hspace{-30pt}\widetilde{G}'''(0)= 6\big(\langle
\psi'',(1-K)^{-1}\psi\rangle\langle \psi',(1-K)^{-1}\psi\rangle\nonumber\\[1ex]
&&\hspace{20pt} -\langle
\psi'',(1-K)^{-1}\psi'\rangle\langle \psi,(1-K)^{-1}\psi\rangle\big) \,.
\end{eqnarray}
Hence, using that
\begin{equation}\label{14a}
 \int^\infty_0
 \lambda^3
(\mathrm{e}^\lambda-1)^{-1}\mathrm{d}\lambda = \pi^4/15\,,
\end{equation}
it holds
\begin{eqnarray}\label{15}
&&\hspace{-46pt} \widetilde{g}_t= 2\gamma^{-4}_t ( \pi^4/15)F
\big(\langle \psi'',(1-K)^{-1}\psi\rangle\langle
\psi',(1-K)^{-1}\psi\rangle\nonumber\\[1ex]
&&\hspace{60pt} -\langle
\psi'',(1-K)^{-1}\psi'\rangle\langle
\psi,(1-K)^{-1}\psi\rangle\big)\,.
\end{eqnarray}

Since $\widetilde{g}_t$ is of order $\gamma^{-4}_t$, in principle,
the terms of $\mathcal{O}(C^2_t)$ could yield a more slowly decaying
contribution. By expanding to second order, one checks that
this contribution is at least of order $\gamma^{-5}_t$. Thus
(\ref{15}) is the leading order in the $1/t$ expansion.\medskip\\
\textbf{Link to the Tracy-Widom distribution.} For a numerical
computation of $\widetilde{g}_t$ it is convenient to use the well
tabulated $F$ \cite{Pra}. One introduces, see TW,
\begin{eqnarray}\label{15a}
&&\hspace{1pt}Q=(1-K)^{-1}\psi\,,\quad P=(1-K)^{-1}\psi'\,,\nonumber\\
&&\hspace{-10pt} q(s)=Q(s)\,,\; p(s)=P(s)\,,\; u(s)=\langle
\psi,Q\rangle\,,\; v(s)=\langle \psi,P\rangle \,.
\end{eqnarray}
To compute (\ref{15}) one still needs $\langle \psi'',Q\rangle$ and
$\langle \psi'',P\rangle$.\medskip\\
\textit{(i) Computation of $\langle \psi'',P\rangle$.} We claim that
\begin{equation}\label{16}
\langle \psi'',P\rangle= \tfrac{1}{2}(v^2-p^2)\,.\medskip
\end{equation}
\textit{Proof}: One starts from TW, below (2.14),
\begin{equation}\label{17}
\langle \psi',P'\rangle=\langle
\psi',(1-K)^{-1}\psi''\rangle
+\langle\psi',[D,(1-K)^{-1}]\psi'\rangle\,.
\end{equation}
On the other hand
\begin{equation}\label{17a}
\langle \psi',P'\rangle= - \psi'(s)P(s)
-\langle\psi'',P\rangle \,.
\end{equation}
Hence
\begin{equation}\label{18}
-2\langle \psi'',(1-K)^{-1}\psi'\rangle=\psi'(s)p+
\langle\psi',[D,(1-K)^{-1}]\psi'\rangle\,.
\end{equation}
From TW (2.13) and the definitions of $R$ and $\rho$ there,
\begin{equation}\label{19}
\langle\psi',[D,(1-K)^{-1}]\psi'\rangle=-v^2
+\langle\psi',R\rangle\langle\psi',\rho\rangle
\end{equation}
and from TW above (2.18),
\begin{equation}\label{20}
\langle\psi',R \rangle=-(\psi'(s)-p)\,,\quad\langle\psi',\rho\rangle=p\,,
\end{equation}
which establishes the claim.{\hspace*{\fill}$\Box$}\medskip\\
\textit{(ii) Computation of $\langle\psi'',Q\rangle$.} We claim that
\begin{equation}\label{21}
\langle\psi'',Q\rangle=-\int^\infty_s p(s')^2 \mathrm{d}s'
+vu-qp\,.\medskip
\end{equation}
\textit{Proof}: Using TW (2.14) one starts from
\begin{eqnarray}\label{22}
&&\hspace{-40pt}\langle \psi',Q'\rangle=-\psi'(s)q-\langle
\psi'',Q\rangle\nonumber\\
&&\hspace{1pt}
=\langle\psi',P\rangle-\langle\psi',Q\rangle u+q\langle R,\psi'\rangle\nonumber\\
&&\hspace{1pt}= \langle \psi',P\rangle -vu+q(p- \psi'(s)) \,.
\end{eqnarray}
Hence
\begin{equation}\label{23}
\langle \psi'',Q\rangle= -\langle \psi',P\rangle +vu-qp\,.
\end{equation}
Next we consider
\begin{eqnarray}\label{24}
&&\hspace{-50pt}\frac{d}{ds}\langle
\psi',P\rangle=-\psi'(s)p-\langle
\psi',R\rangle p\nonumber\\
&&\hspace{6pt}= - \psi'(s)p-(-\psi'(s)+p)p=-p^2 \,,
\end{eqnarray}
which establishes the claim. {\hspace*{\fill}$\Box$}\medskip\\

Using the identities from TW, it holds
\begin{equation}\label{25}
u=F'/F\,,\quad v=F''/2F\,,\quad q^2=u^2-2v\,,\quad p=q'+qu\,.
\end{equation}
In combination with (\ref{15}), (\ref{15a}), (\ref{16}), and (\ref{21}), one arrives
at
\begin{eqnarray}\label{26}
&&\hspace{-45pt}\widetilde{g}_t(s)= (\pi^4/15)\gamma^{-4}_t F(s)
\nonumber\\
&&\hspace{-5pt}\times\Big(-2v(s)\Big(q(s)p(s)+ \int^\infty_s
p(s')^2\mathrm{d}s'\Big)+u(s)\big(v(s)^2+p(s)^2\big)\Big)\,,
\end{eqnarray}
which is the starting point for numerical evaluations. In our notation, \cite{Pra} tabulates $F,q,q',u$. 
From (\ref{25}) we then deduce  $p$ and $v$ and only the computation of the integral remains.

\end{appendix}


\begin{figure}[b]
  \psfrag{u}{$s$}
  \psfrag{g}{}
\begin{center}
 \includegraphics[scale=1.]{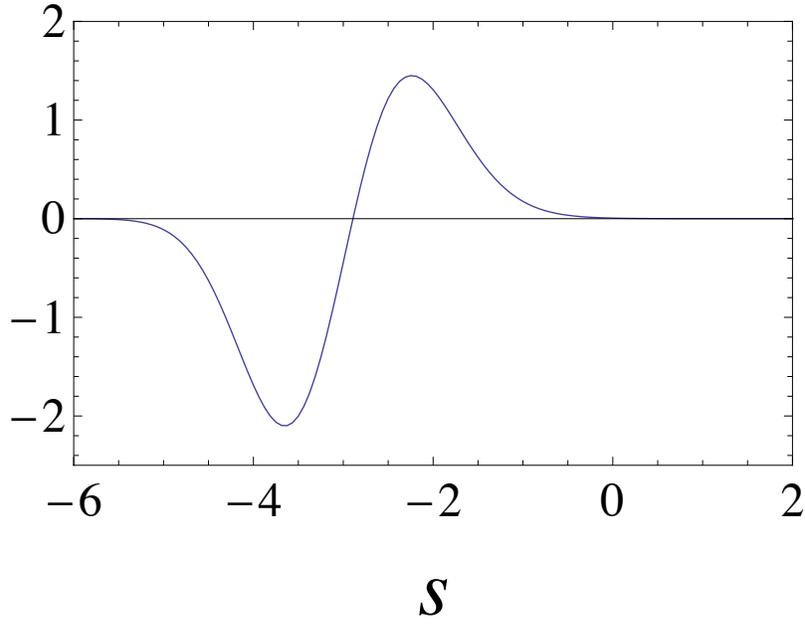}
\end{center}
\caption{
The first order correction $\tilde{g}_t(s)$ in
(\ref{6.2}). The vertical scale is  for $\gamma_t=1$.} 
\end{figure}
\begin{figure}[t]
\begin{center}
\includegraphics[scale=1.]{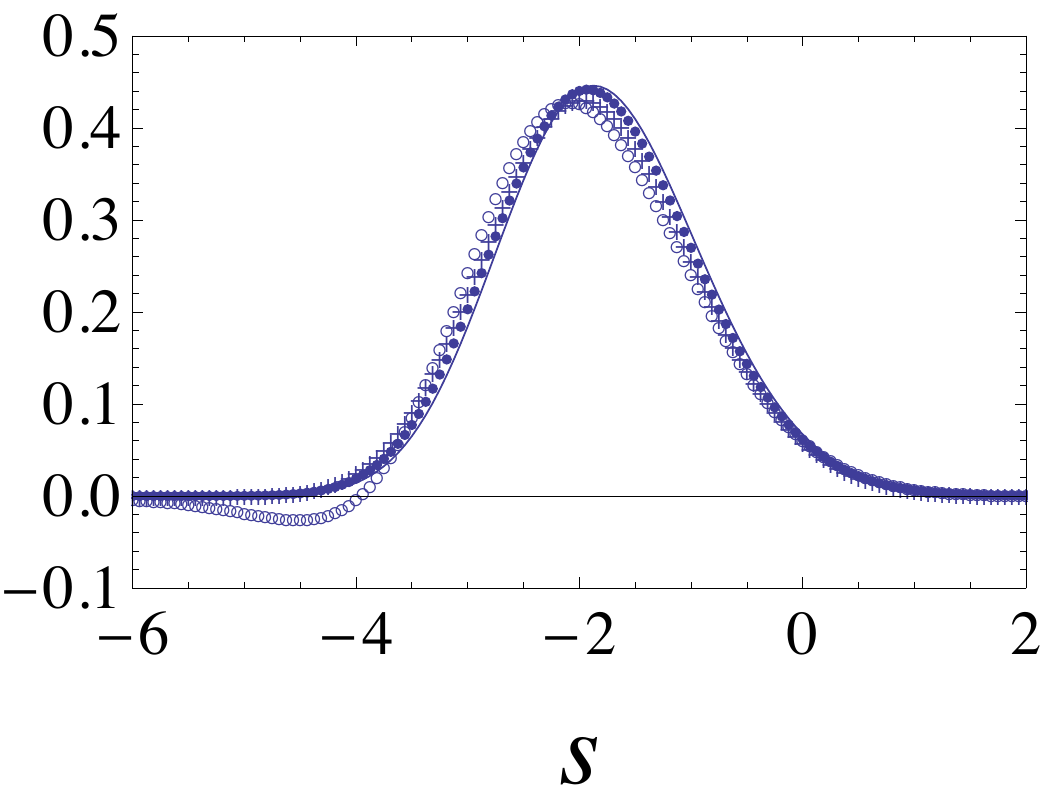}
\end{center}
\caption{The first order approximation $\rho^{(1)}_t (s)$ in
(\ref{6.3}) for $\gamma_t=2\,(\circ),5\,(+),10\,(\bullet)$.} 
\end{figure}
\begin{figure}[!b]
\begin{center}
\includegraphics[scale=1.]{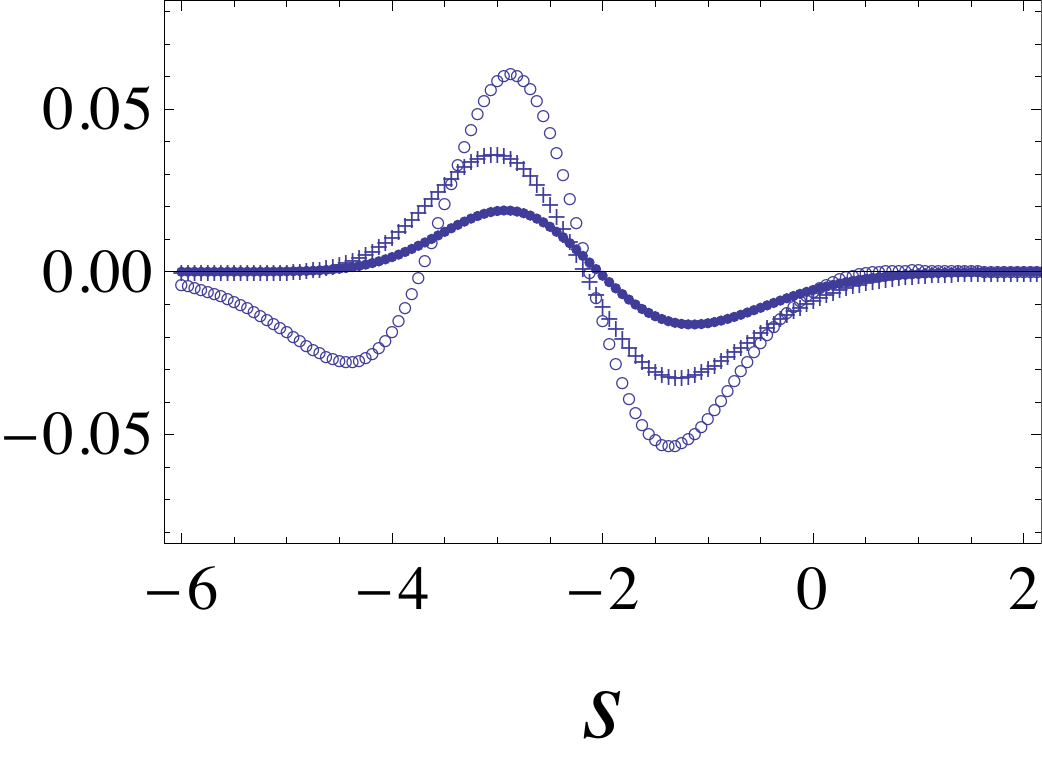}
\end{center}
\caption{The difference $\rho^{(1)}_t (s) -\rho_\mathrm{TW}(s)$ for $\gamma_t=2\,(\circ),5\,(+),10\,(\bullet)$.} 
\end{figure}
\begin{figure}[h]
\begin{center}
\includegraphics[scale=1.]{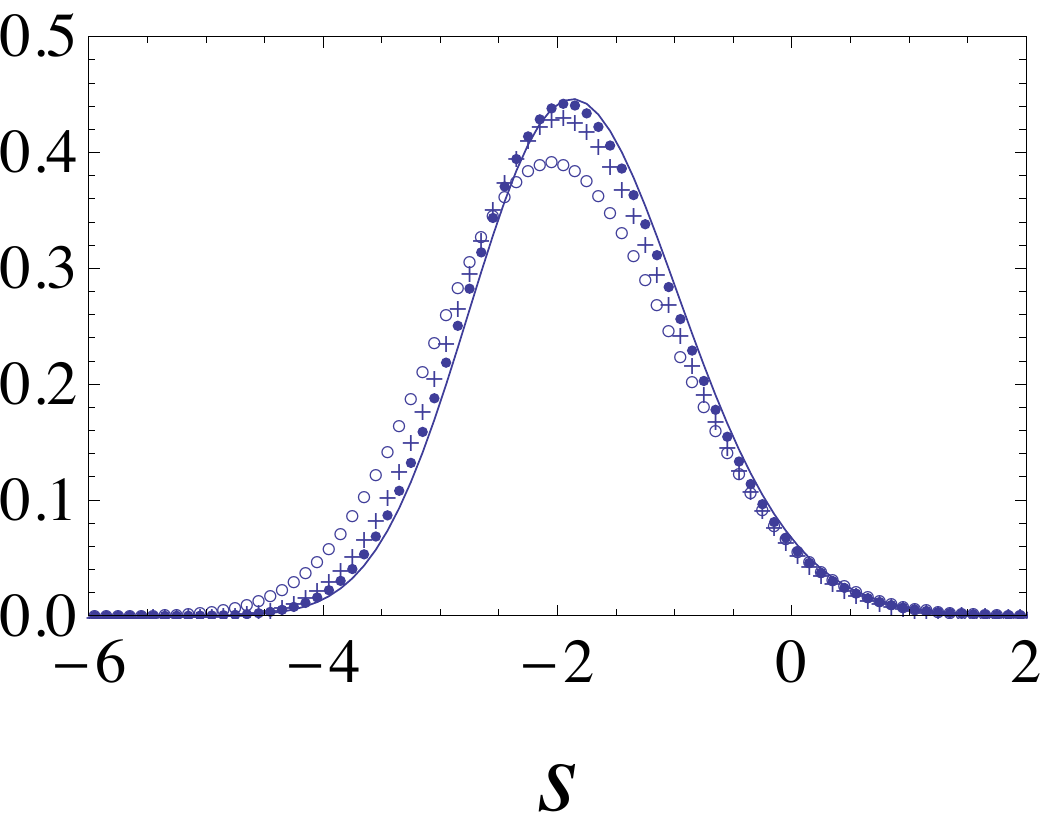}
\end{center}
\caption{KPZ probability densities $\rho_t (s)$ for $\gamma_t=2\,(\circ),5\,(+),10\,(\bullet)$ based on the evaluation of the determinants.} 
\end{figure}
\begin{figure}[h]
\begin{center}
\includegraphics[scale=1.]{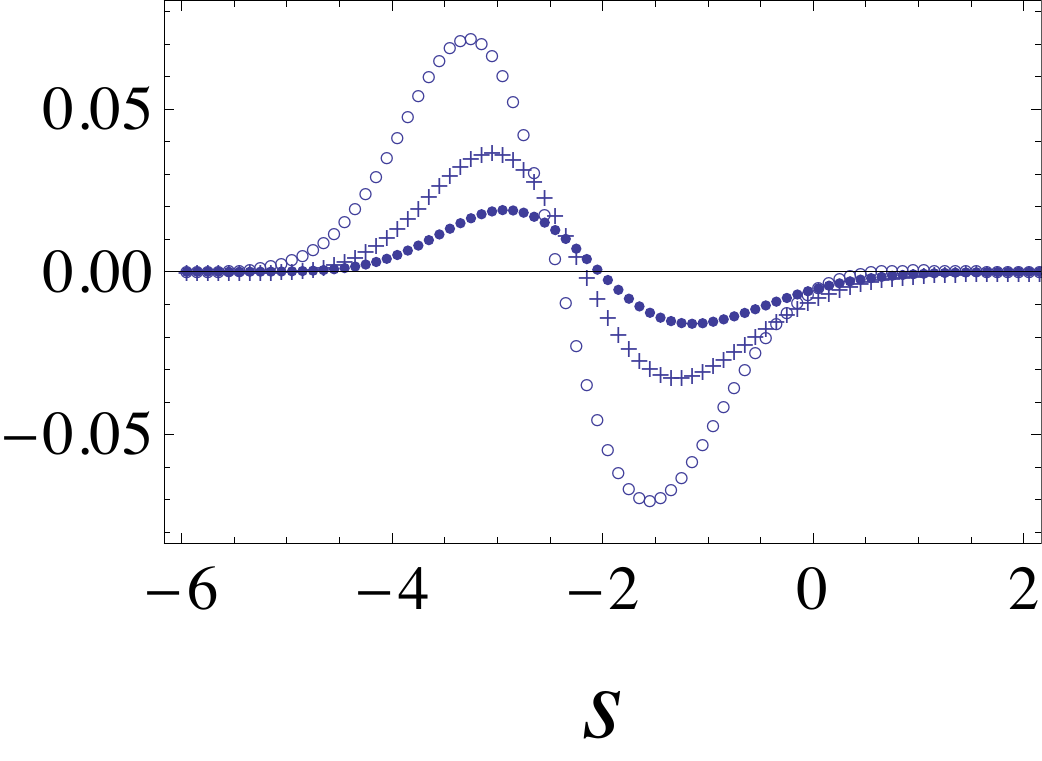}
\end{center}
\caption{The difference $\rho_t (s)  -\rho_\mathrm{TW}(s)$ for $\gamma_t=2\,(\circ),5\,(+),10\,(\bullet)$
based on the evaluation of the determinants.} 
\end{figure}
\begin{figure}[!h]
\begin{center}
\includegraphics[scale=1.]{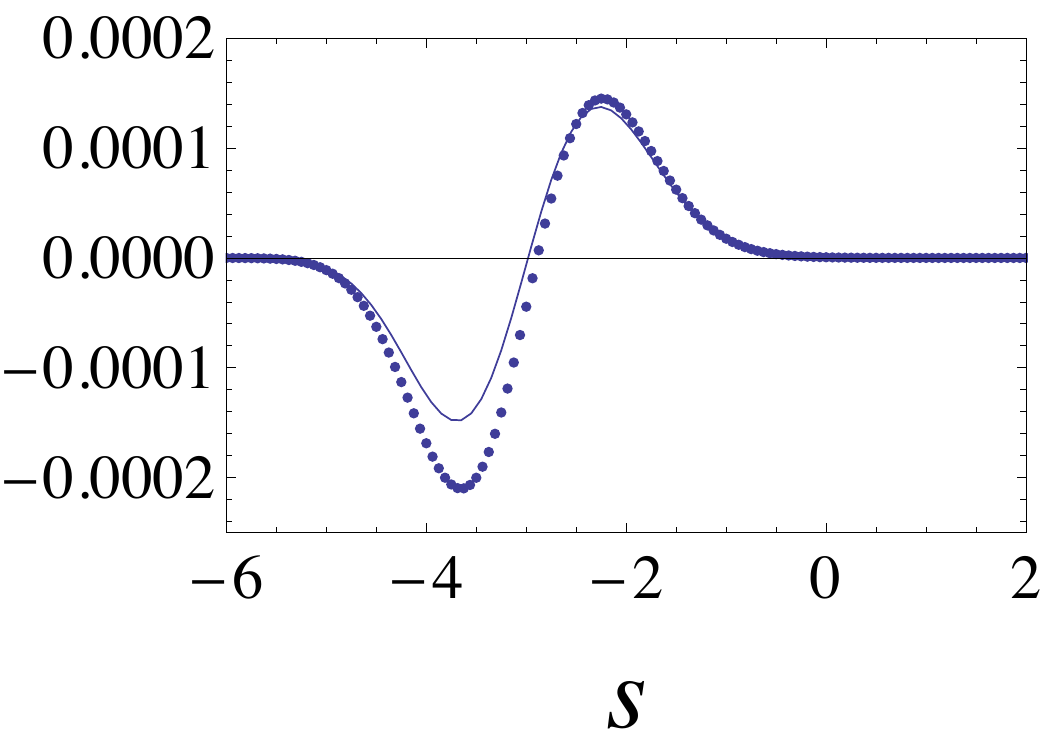}
\caption{$\tilde{g}_t(s)$ ($\frac{\hspace{12pt}}{\hspace{12pt}}$) and $g_t(s)- \rho_\mathrm{TW}(s)$ ($\bullet$) for $\gamma_t =10$.} 
\end{center}
\end{figure}
\begin{figure}[h]
\begin{center}
\includegraphics[scale=1.]{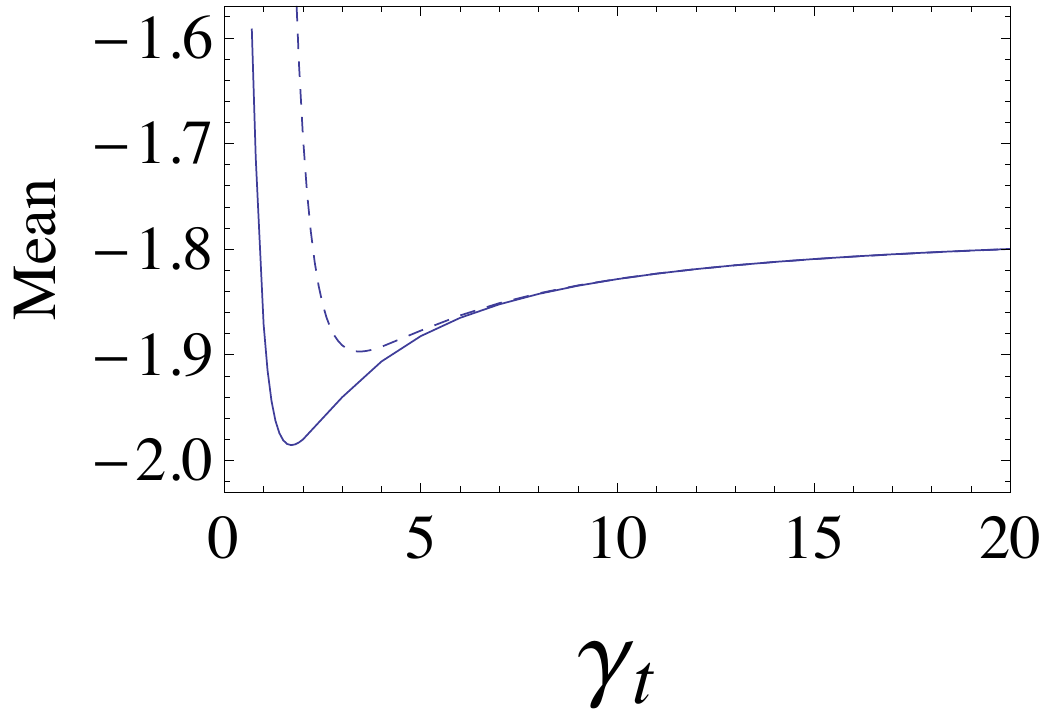}
\caption{First moment of  $\rho_t $ ($\frac{\hspace{12pt}}{\hspace{12pt}}$) and $\rho^{(1)}_t $ 
($\frac{\hspace{3pt}}{\hspace{3pt}}$$\frac{\hspace{3pt}}{\hspace{3pt}}$$\frac{\hspace{3pt}}{\hspace{3pt}}$) as a function of $\gamma_t$.} 
\end{center}
\end{figure}
\begin{figure}[!b]
\begin{center}
\includegraphics[scale=.9]{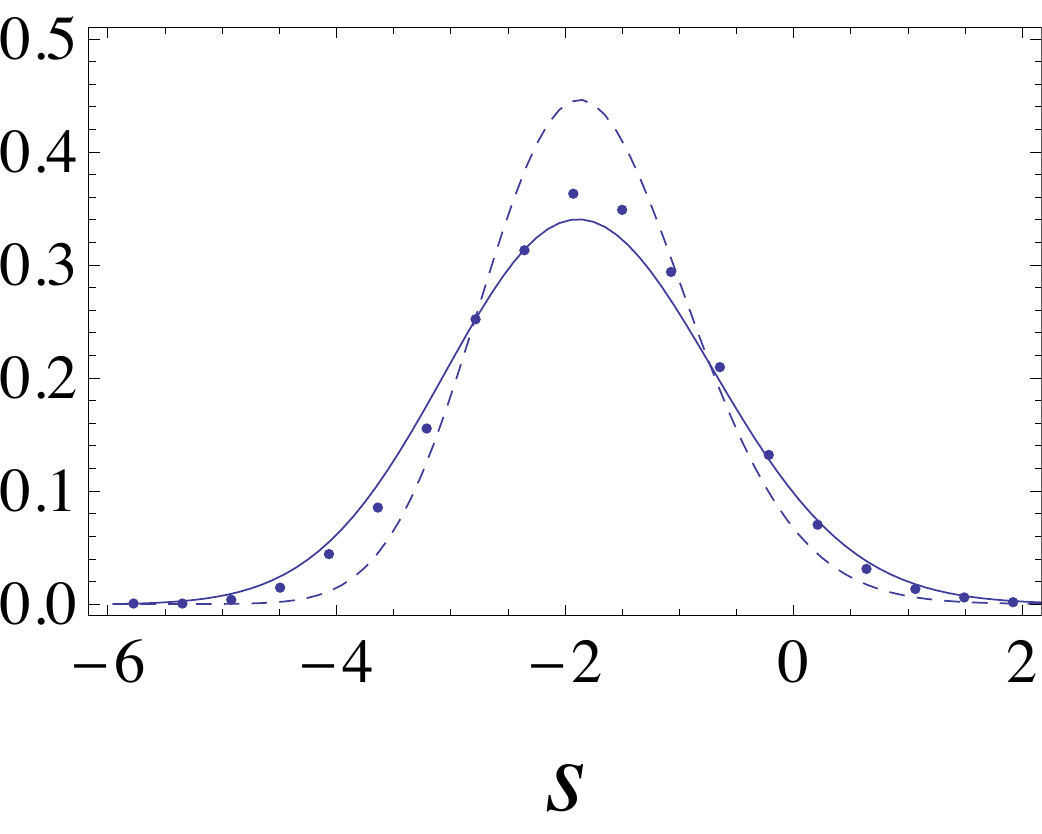}
\caption{The exact KPZ density at $\gamma_t = 0.94$ ($\frac{\hspace{12pt}}{\hspace{12pt}}$) and the 
PASEP Monte Carlo at $q=0.6$, $t = 1024$ MC steps ($\bullet$). The dashed line is the Tracy-Widom density.} 
\end{center}
\end{figure}

\end{document}